\documentclass[12pt]
{iopart}
\usepackage{graphicx}
\usepackage{color}
\begin{document}
\title[An autoencoder neural network integrated into gravitational-wave burst searches]{An autoencoder neural network integrated into gravitational-wave burst searches to improve the rejection of noise transients}
\author{Sophie Bini$^{1,2}$, Gabriele Vedovato$^3$, Marco Drago$^{4,5}$ Francesco Salemi$^{1,2}$, Giovanni A. Prodi$^{2,6}$}

\address{$^1$ Universit\`a di Trento, Dipartimento di Fisica, I-38123 Povo, Trento, Italy}
\address{$^2$ INFN, Trento Institute for Fundamental Physics and Applications, I-38123 Povo, Trento, Italy}
\address{$^3$ INFN, Sezione di Padova, I-35131 Padova, Italy}
\address{$^4$ Universit\`a di Roma  La Sapienza, I-00185 Roma, Italy}
\address{$^5$ INFN, Sezione di Roma, I-00185 Roma, Italy}
\address{$^6$ Universit\`a di Trento, Dipartimento di Matematica, I-38123 Povo, Trento, Italy, INFN}
\ead{sophie.bini@unitn.it}

\begin{abstract}
The gravitational-wave (GW) detector data are affected by short-lived instrumental or terrestrial transients, called “glitches”, 
which can simulate GW signals. Mitigation of glitches is particularly difficult for algorithms which target 
generic sources of short-duration GW transients (GWT), and do not rely on GW waveform models to distinguish astrophysical signals from noise, such as Coherent WaveBurst (cWB). 
This work is part of the long-term effort to mitigate transient noises in cWB, which led to the introduction of specific estimators, and a machine-learning based signal-noise classification algorithm. Here, we propose an autoencoder neural network, integrated into cWB, that learns transient noises morphologies from GW time-series. We test its performance on the glitch family known as “blip”. 
The resulting sensitivity to generic GWT and binary black hole mergers significantly improves when tested on LIGO detectors data from the last observation period (O3b). At false alarm rate of one event per 50 years the sensitivity volume increases up to 30\% for signal morphologies similar to blip glitches. 
 In perspective, this tool can adapt to classify different transient noise classes that may affect future observing runs, enhancing GWT searches.
\end{abstract}
\noindent{\it Keywords\/}:{ gravitational-wave bursts, autoencoder neural network, transient noises}
\maketitle

\section{Introduction}
Gravitational-wave (GW) interferometers detected 90 astrophysical signals originating from the coalescence of compact objects, mainly black holes (BH) but also neutron stars \cite{GWTC1,GWTC2, GWTC3}. These exceptional observations have been achieved thanks to Advanced LIGO detectors \cite{TheLIGOScientific:2014jea} and Advanced Virgo \cite{TheVirgo:2014hva}, that together with KAGRA \cite{Aso:2013eba} and GEO600 \cite{willke2002geo} constitute the global network of GW detectors of the LIGO-Virgo-KAGRA (LVK) collaboration.\\
One of the major challenges for both detector and data-analysis experts is represented by short-duration disturbances, usually referred to as glitches, that are present in GW data with both high signal-to-noise ratio (SNR) and high rate.  Short-duration noises are particularly concerning because they can mimic GWTs originated for example by the coalescence of compact binaries, especially in the case of high mass binary black hole (BBH) mergers, or other sources still not-detected as supernovae \cite{abbott2020optically}, isolated neutron stars \cite{lopez2022search, abbott2022search_mag}, cosmic strings \cite{abbott2021constraints} GW non-linear memory \cite{ebersold2020search} and radiation-driven BBH
capture events \cite{ebersold2022observational}.
Moreover, transient noises might overlap with true astrophysical signals and affect the estimation of the parameters of the GW signals \cite{pankow2018mitigation}, their sky localization \cite{macas2022impact} and the studies performed to test the General Relativity \cite{kwok2022investigation}. 
These transients noises are caused by the instrument itself or by its interaction with the environment  \cite{davis2021ligo, acernese2022virgo}. 
Ideally, the best strategy to reduce their impact is to track back their origin and remove the causes \cite{schofield_glitch}. 
When it is not possible to resolve the root cause, but the coupling between the noise source and the detector output is known and reproducible, periods of data on the order of seconds or hours can be excluded. On the other side, if the coupling has not completely been identified, the transient noises cannot be vetoed safely, 
and data analysis algorithms design specific methodologies to minimize their impact (see Section \ref{sec:Blip} and Appendix \ref{sec:Qveto}).\\
In this work we focus on model-agnostic algorithms, which do not assume any GW templates and are open-wide to generic GWTs in multiple detectors data, and so are more affected by transient noises. 
We introduce a deep-learning algorithm to classify specific glitch morphologies and mitigate their impact in generic GWT searches. The algorithm proposed consists of a neural network architecture, called \textit{autoencoder}, that learns the morphology of a specific class of glitch from the time-series. When it is applied to a generic detected event, it estimates the similarity between the tested event and the noise class learned. We test this methodology on blip glitches, the  family which most affected the GWT search for Advanced detectors \cite{GWTC1, davis2021ligo}. 
A crucial aspect of this methodology is that the network is trained only on examples of transient noises present in the detector data, and no GW signal model needs to be considered. This requirement is determined by the desire to apply the proposed method to generic GWT searches.\\
Here, we implement the autoencoder neural network on Coherent WaveBurst (cWB) \cite{klimenko2008coherent, drago2021coherent}, a weakly—modelled algorithm used also by LVK collaboration for generic GWT detection and reconstruction \cite{Abadie:2012rq, Abadie:2010mt, Abbott:2009zi, Abbott:2005at, Abbott:2005rt, Abbott:2005fb, Abbott:2004rt, O3allskyshort}. 
Over the years, different strategies to mitigate the impact of transient noises have been integrated into cWB: two estimators, called $Qveto_0$ and $Qveto_1$, have been designed to pinpoint short-duration glitches, and 
recently cWB has been enhanced by a signal-noise classification with the decision-tree learning algorithm XGBoost \cite{mishra2021optimization}. This latter methodology exploits a set of summary statistics computed by cWB, and has shown to increase the search sensitivity for compact binary coalescence and generic GW searches \cite{mishra2021optimization, mishra2022search, szczepanczyk2022all}. Here, we propose an additional estimator, computed using the autoencoder neural network, that can be straightforwardly included in the statistics used to build the XGBoost model, further enhancing the discrimination of glitches.\\
The content of the following sections is the following: Section \ref{sec:Blip} describes in more detail short-duration glitches and the investigation performed to find their origins, Section \ref{sec:cWB} introduces cWB and reviews the strategy adopted so far to mitigate transient noises. 
Section \ref{sec:ae} outlines the autoencoder neural network, its architecture, and the training and test datasets. The results obtained with the introduction of this methodology on ad-hoc waveforms and on BBH simulations are reported in Section \ref{sec:results}.

\section{Transient noises in gravitational-wave data}
\label{sec:Blip}
During the third observing run (O3), which took place from April 2019 to March 2020, the median rate of glitches with SNR$>$6.5 was about 0.3 min$^{-1}$ in LIGO Hanford, 1 min$^{-1}$ in LIGO Livingston and 0.8 min$^{-1}$ in Virgo, with an increase from November 2019 to January 2020 due to adverse weather conditions \cite{GWTC2, GWTC3}.\\
Several methods based on machine-learning techniques have been developed in the latest years to characterize and mitigate transient noises \cite{Cuoco_2020}.
Many studies propose glitch classification into families according to their morphology in time-series \cite{mukund2017transient,powell2015classification} or in time-frequency representations \cite{george2018classification,razzano2018image}. Classification is crucial to characterize transient noises and identify their root causes: it allows ranking by quantity and characteristics, improving the production of specific vetoes. 
A successful citizen-science project for supervised classification of GW transient noises is GravitySpy~\cite{zevin2017gravity}
, 
which couples a neural network together with human classification performed by citizen scientists. Recently, it has been joined by the project GWitchHunter, more oriented to Virgo glitches \cite{RAZZANO2023167959}.\\ 
These studies indicate blip glitches as one of the most concerning classes \cite{cabero2019blip,abbott2016characterization, abbott2018effects, glanzer2022data}: typically, they have a sub-second duration $\mathcal{O}(10)$ms and a large frequency bandwidth $\mathcal{O}(100)$Hz. 
Their rate during the second observing run (O2) was typically of 2 per hour in LIGO, increased to 4 per hour in LIGO Livingston during the third observing run (O3) \cite{davis2021ligo}.
Blip glitches appear to have multiple subclasses that might have different origins.
Despite automated algorithms that correlate the detector output with the auxiliary channels, used to monitor the state of the instruments and their environment \cite{smith2011hierarchical}, their origin is still largely unknown. 
Four subsets of blip have been found correlated respectively with humidity, laser intensity stabilization, computer errors and power recycling cavity controls, but these correlations regard a minority of the total number of classified blip ($\sim 8\%$ in LIGO Hanford and $\sim 2\%$ in LIGO Livingston during the first and second observing runs) \cite{cabero2019blip}. Possible correlations with cosmic rays or errors in the data acquisition system have also been investigated, but no evidence was found \cite{davis2021ligo}.\\
Blip glitches are responsible for a major fraction of the unvetoed high SNR triggers, reducing the effectiveness of GW searches. Thus, specific techniques have to be adopted by each GW data analysis algorithm. 
For example, the template-based algorithm PyCBC \cite{usman2016pycbc} has implemented a specific methodology to mitigate the impact of blip glitches on the search for high mass BBH mergers: 
while the GW models for these sources might not be distinguishable from blip at first sight, blips actually have an excess of power at middle to high frequencies, which is not belonging to the GW template. To highlight this discrepancy, a consistency test between the high mass BBH waveforms and the transient noises has been introduced  in PyCBC \cite{nitz2018distinguishing}, leading to a significant improvement in the algorithm performance.\\

\begin{figure*}[ht]
\begin{minipage}[b]{0.47\linewidth}
\centering
\includegraphics[width=0.9\textwidth]{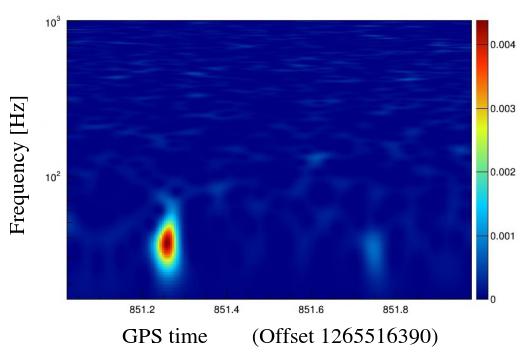}
\end{minipage}
\hspace{0.5cm}
\begin{minipage}[b]{0.47\linewidth}
\centering
\includegraphics[width=0.9\textwidth]{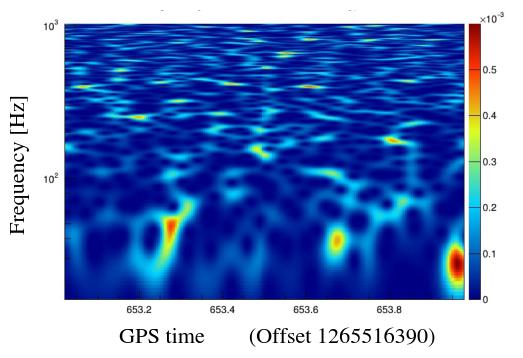}
\end{minipage}
\begin{minipage}[b]{0.47\linewidth}
\centering
\includegraphics[width=0.9\textwidth]{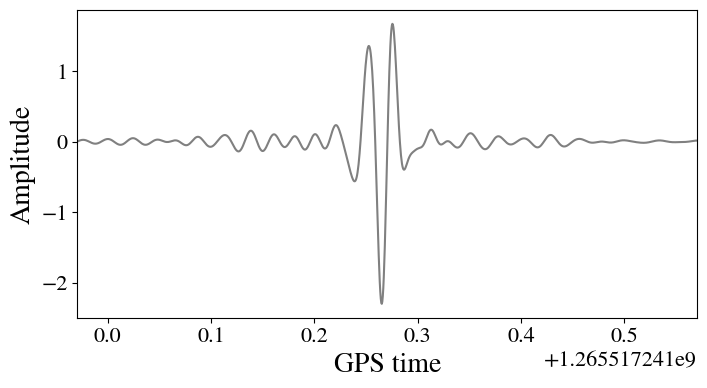}
\end{minipage}
\hspace{0.5cm}
\begin{minipage}[b]{0.47\linewidth}
\centering
\includegraphics[width=0.9\textwidth]{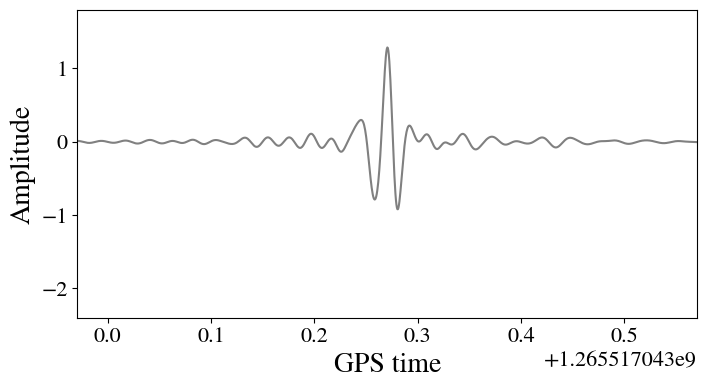}
\end{minipage}
\caption{Example of the spectrograms (top) and respective time-series (bottom) of a noise event detected by cWB in LIGO time-shifted data (see Section \ref{sec:cWB}). The spectrogram on the left (LIGO Livingston) shows clearly a blip glitch with reconstructed $SNR=12$. On the right (LIGO Hanford) the spectrogram shows a weaker disturbance ($SNR=6$) with multiple low frequency spots (note that the colour scales are different for the two spectrograms). In both cases, the reconstructed time-series show a blip-like morphology. 
}
\label{fig:blip_example}
\end{figure*}
In the next sections, we discuss the strategies implemented in cWB to mitigate the impact of transient noises, and we propose a deep-learning algorithm which targets blip glitches.
\section{Transient noises mitigation in coherent WaveBurst}
\label{sec:cWB}
CWB is an algorithm widely used by the LVK collaboration for detection and reconstruction of generic GWT, including signals from compact binary coalescences. cWB combines coherently the detector network time-frequency maps, computed with multi-resolution Wilson-Daubechies-Meyer wavelet transform~\cite{necula2012transient}, and maximizes a likelihood ratio statistic over all sky directions.  
To characterize each detected event, several summary statistics are estimated, as the coherence across the detector network, the signal strength, the peak frequency and the duration. cWB also provides a reconstructed  waveform of the detected events for each detector.\\ 
To reject the events arising from non-stationary detector noise, cWB computes the coherent energy $E_C$ in the detector network and the residual noise energy $E_N$, estimated subtracting the reconstructed waveform from the data \cite{klimenko2008coherent}. The events with network correlation coefficient $cc = E_C /(E_C + E_N)$ below a certain threshold are rejected. However, due to the high rate of single detector glitches, there is a non-negligible probability of having accidental coincidences in multiple detectors between independent transient noises. Moreover, glitches which occur in a single detector could match part of the noise in the other detectors, especially in the case of glitches with a simple morphology, as blips. An example of the reconstruction provided by cWB of a noise event is reported in Fig. \ref{fig:blip_example}.\\
For these reasons,  model-agnostic methods are strongly affected by short-duration disturbances and  the requirement on the correlation coefficient $cc$ is not enough to mitigate transient noises. To reduce the false alarm rate an extra step is necessary. In cWB a separation between GW signal and noise was implemented based on two statistics, referred to as $Qveto_0$ and $Qveto_1$, computed from the reconstructed waveforms (Appendix \ref{sec:Qveto}).\\
Recently, to automate the signal-noise separation and avoid the application of hard thresholds, 
\label{sec:xgb}
it has been implemented a procedure based on a decision tree learning algorithm, called XGBoost \cite{chen2015xgboost}, which performs a binary classification between GW signals and noise \cite{mishra2021optimization, mishra2022search, szczepanczyk2022all}. In order to learn the differences between the population of the signal and of the noise, XGBoost receives in input a list of eight cWB summary statistics that do not depend on the waveform morphology. 
The signal population is modeled using generic White Noise Burst (WNB) waveforms, which are basically random noise constrained in a certain time-frequency range sampled from a random distribution. To represent the noise population we employ 
the time-shift analysis: the data of one detector is shifted with respect to the other detector so that the coincident events by construction do not have an astrophysical origin, but they are solely due to non-stationary detector noises. The cWB events detected on time-shifted data are referred to as \textit{background} events.
The XGBoost model computes a penalty factor which ranges from 0 for noise and 1 for signal, and it updates the cWB ranking statistic $\rho$ \cite{mishra2022search}. 
%
%
%

\section{Further discrimination of transient noises by an autoencoder neural network}
\label{sec:ae}
We implement an autoencoder neural network to classify transient noises from cWB reconstructed waveforms. 
An autoencoder is an unsupervised learning neural network that compresses the input data into a lower dimensional space, called \textit{latent space}, and then re-constructs an output with the original dimension. Performing the compression, the autoencoder highlights the presence of structures in the input data, and disregards redundancies in the data.
The autoencoder architecture is used in GW physics for features extraction, data denoising and anomaly detection  (\cite{gabbard2022bayesian, ormiston2020noise, corizzo2020scalable,morawski2021anomaly, shen2019denoising}. \\
Here, the autoencoder performs an anomaly detection task: 
the training dataset contains time-series examples belonging to a single class, so that the network learns that specific morphology. Once the autoencoder is applied to a time-series with a different signature, the \textit{anomaly}, the network struggles to reconstruct it properly. The goodness of the reconstruction is evaluated through the mean square error (MSE) between the input time-series ($X_{i, input}$) and the one reconstructed by the autoencoder ($X_{i, ae}$):
\begin{equation}
    MSE = \sum_{i=0}^n{(X_{i, input} - X_{i, ae})^2}
    \label{eq:MSE}
\end{equation}
where $n$ is the time-series length. The MSE is computed for each detector, and when searching for GWTs it is weighted according to the SNR square of the event in each detector. In the following, the weighted MSE estimator will be referred to as the autoencoder statistics.

\subsection{The architecture}
An autoencoder network learns the key features of the input data, which in this case is a time-series $x_i$ with $n$ data points. 
The network consists of two parts: an encoder $f_E(\cdot)$ that compresses the input representation into a lower dimension latent space, and a decoder $g_D(\cdot)$
that converts the latent representation to the original format. The encoder and decoder weights, $E$ and $D$, are found by minimizing the difference between the input $x_i$ and the autoencoder output $g_D(f_E(x_i))$. The error function $J$ is the mean square errors between the input and the output:
\begin{equation}
J(E, D ; x_i ) = \frac{1}{n} \sum_{i=1}^{n} || g_D(f_E(x_i))-x_i||^2
\label{eq:loss}
\end{equation}
At the beginning of the training procedure the weights are set randomly, then they are updated minimizing Eq.\ref{eq:loss} for each $x_i$ present in the training dataset. In other words, the weights are updated so that $g_D(f_E(x_i)) \sim x_i$: the network searches for an approximation to the identity function but, as the latent space has a lower dimension than the input data, the algorithm is forced to learn a compressed representation. 
The autoencoder performs a dimensional reduction, but differently from common tools as Principal Component Analysis \cite{bro2014principal}, can learn non-linear and complex features.\\
Once trained, the weights $E$ and $D$ are fixed and the network is able to reconstruct properly a time-series only if it has a morphology similar to the examples present in the training dataset. The goodness of the reconstruction is estimated with the MSE (Eq.\ref{eq:MSE}).\\
We build the network using the deep-learning application programming interface Keras \cite{chollet2015keras}. The autoencoder is made of multiple convolutional layers, one of the most recognized machine-learning algorithm to extrapolate relevant features in time-series and images. More details on the architecture of the network are reported in the Appendix \ref{appendix}. \\ 
A crucial aspect of the architecture is the latent space dimension: the lower the dimension, the stronger the compression of the input, and the less information is retained by the network. Test at different compression factors, showed that for higher compression the network learns only loud blip-like morphology, and it struggles when the combination of the glitch with the detector noise leads to slightly different morphology, resulting in a poor ability to mitigate the background events.
\subsection{The autoencoder input data}
\label{sec:input}
The inputs $x_i$ of the autoencoder are cWB reconstructed time-series which, thanks to the whitening procedure\footnote{\textit{https://gwburst.gitlab.io/documentation/latest/html/faq.html\#the-whitening}} and the selection of the most energetic wavelets in the time-frequency maps, appear much cleaner than the raw GW detector data. Before being processed by the autoencoder, each cWB reconstructed time-series, sampled at 2048 Hz, is windowed to 416 data points (corresponding to $\sim$0.2 s), and centred around the absolute maximum value. Several window lengths have been tested from about 200 to 800 data points, but this choice is a good compromise between containing the entire blip-like evolution and minimizing the information to be learnt. Next, the time-series is normalized in amplitude between $[0,1]$, as suitable for neural networks. Some examples of cWB reconstructed waveforms processed to be input to the autoencoder are shown in Fig. \ref{fig:timeseries}.

\subsection{The training dataset}
\begin{figure}[ht]
    \centering
	\includegraphics[width=0.65\columnwidth]{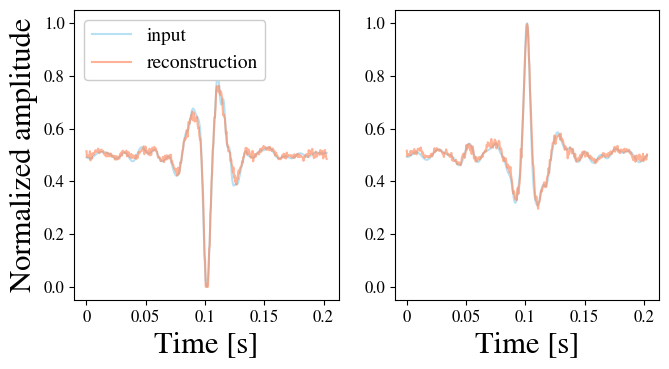}
	\caption{Two examples of blip time-series according to the GravitySpy classification detected by cWB in LIGO Hanford. In blue the autoencoder inputs $x_i$, that are cWB reconstructed waveforms windowed and normalized as described in Section \ref{sec:input}). In orange the autoencoder reconstructions $g_D(f_E(x_i))$.}
	\label{fig:timeseries}
\end{figure}

The training dataset is made of blip glitches according to the GravitySpy \cite{zevin2017gravity, glanzer2022data} classification. To retrieve the blip time-series as observed by cWB, we run the pipeline in each single detector, obtaining a list of detected events, glitches and eventually GW signals, in each detector. Next, we select blip glitches comparing the GPS time of cWB single detector events and GravitySpy blip GPS time. We employ two similar glitch families classified by GravitySpy, nominally \textit{blip} and \textit{tomte}.\\
Deep-learning methods benefit from the largest possible training dataset, and in order to collect more training samples several strategies have been developed, usually referred to as \textit{data augmentation} techniques. In this work, we double the amount of the training dataset including also the vertical flip of each time-series. 
Moreover, we add gaussian noise to each input data to improve the capability of the autoencoder network to pinpoint also low SNR glitches. The resulting training dataset is composed of 4608 glitches occurred during the second period of the third observing run (O3b). 
Additional 512 classified samples constitute the validation dataset, which tests that the neural network is not over-fitting. We retrieve blip glitches from the two LIGO detectors, but there is evidence for the presence of such glitches also in Virgo detector \cite{cabero2019blip}, whose transient noises are also classified by GravitySpy. Fig. \ref{fig:timeseries} shows two examples of time-series present in the validation dataset and the reconstruction achieved by the autoencoder. 
\subsection{Signal models for testing}
\label{sec:test}
We test the autoencoder neural network on three sets of waveforms. The first is made of ad-hoc signals, generally used in generic GWT searches to evaluate the sensitivity of a weakly modelled pipeline over a wide parameter space \cite{O3allskyshort}. They include Sine-Gaussian (SG), Gaussian Pulse (GA) and White Noise Burst (WNB). The SG waveforms are characterized by the central frequency, and the quality factor $Q$. The GA signals are characterized by the duration, and the WNB by their bandwidth, duration and lower frequency bound. \\
The second set of waveforms contains cosmic strings simulations, potential burst sources supposed to originate after a spontaneous phase transition in the early Universe \cite{abbott2021constraints}. Here, we inject cosmic strings from cusps \cite{damour2005gravitational}, which are characterized by the amplitude, and frequency range of [1 Hz, 1500 Hz]. These waveforms have been included to test the robustness of the proposed methodology on potential GW signals with a morphology similar to blip glitches. The results achieved cannot be compared directly to the LVK search \cite{abbott2021constraints}, because the detection efficiency is parametrized differently.\\
The third set of waveforms is composed of binary black hole (BBH) coalescences with quasi circular orbits, as they represent the  GW signals mainly observed so far. Their waveforms include both precession and higher order modes, and they are computed using \texttt{SEOBNRv4PHM} model \cite{ossokine2020multipolar}. 
\section{Results}
\label{sec:results}
We compute the autoencoder statistic, defined in Eq.\ref{eq:MSE} as the MSE between the cWB reconstructed waveform of each detected event and the corresponding time-series reconstructed by the autoencoder, for the background events and the injected signals.  A low MSE suggests that the event considered has a blip-like signature, while a high MSE indicates a different morphology.
We include the obtained MSE in the list of summary statistics used by XGBoost to perform the signal-noise separation in cWB. This configuration will be referred as \texttt{XGBoost + AE} model, while the configuration without the autoencoder statistic will be referred simply as \texttt{XGBoost} model. The cWB pipeline's set up and the hyperparameters employed for the XGBoost tuning are equal for the two configurations, and the same used for the generic GWT search performed with cWB enhanced by XGBoost, which provides the most stringent constraints on the isotropic emission of GW energy from burst sources to date \cite{szczepanczyk2022all}. Using these two configurations, we analyse 40 days, between February and March 2020, of coincident data between the LIGO detectors. We accumulate about 380 years of background using the time-shift analysis (Section \ref{sec:xgb}), $70\%$ of which is used to train the XGBoost model and the remaining $30\%$ for testing. Similar results are obtained also with different percentages dedicated to the training and the testing.\\ 
First, we discuss the impact of the autoencoder statistic on the background events distribution (Section \ref{sec:results_bkg}), then we report the search sensitivity achieved on ad-hoc waveforms, cosmic strings, and BBH simulations (Section \ref{sec:result_sim}).
\subsection{Background events distribution}
\label{sec:results_bkg}
The background events distribution is crucial to assess the significance of the detected events in terms of the inverse false alarm rate (IFAR), where IFAR=1/FAR. The FAR of a detected event with ranking statistic $\rho_k$ is:
\begin{equation}
    FAR = \frac{N(\rho_k)}{T}
\end{equation}
where $N(\rho_k)$ is the number of background events with $\rho > \rho_k$ and $T$ the accumulated background time. 
\begin{figure*}[ht]
\begin{minipage}[b]{0.49\linewidth}
\centering
\includegraphics[width=0.99\textwidth]{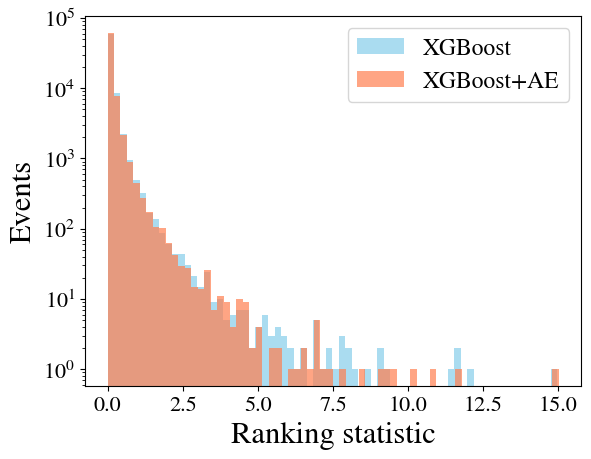}
\end{minipage}
\hspace{0.5cm}
\begin{minipage}[b]{0.49\linewidth}
\centering
\includegraphics[width=0.99\textwidth]{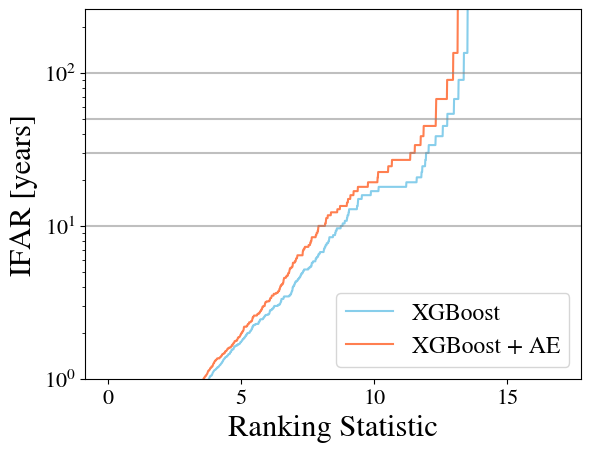}
\end{minipage}
\caption{(Left) Background events versus the ranking statistic $\rho$ for the \texttt{XGBoost} model (blue) and the \texttt{XGBoost + AE} one (orange). (Right) IFAR versus the ranking statistic for the background events for the two models. The \texttt{XGBoost + AE} model reduces the number of background events at $\rho > 5$, so that at a fixed IFAR threshold the corresponding $\rho$ is lower. The gray lines mark the IFAR thresholds for which the search sensitivity is reported in Fig. \ref{fig:volume} and Fig. \ref{fig:ifar_V}.} 
\label{fig:bkg}
\end{figure*}
So, the fewer the background events and the lower their $\rho$, the more sensitive the search for GWT will be. In Fig.\ref{fig:bkg} (left) the background events are shown versus the cWB ranking statistic $\rho$. The autoencoder further mitigates the background distribution, with 28 events at  $\rho >5$ using the \texttt{XGBoost + AE} model, rather than 47 events with the \texttt{XGBoost} model. 
This enhancement can be appreciated in  Fig.\ref{fig:bkg} (right), which shows the IFAR versus the ranking statistics for the background distribution for the two models considered. At a fixed $\rho$, the inclusion of the autoencoder's statistic increases the corresponding IFAR, meaning that a potential GW signal detected by cWB with a certain $\rho$ can have a higher significance with the \texttt{XGBoost + AE} model, or in other words that using the autoencoder we can assing the same IFAR to weaker GW signals.
\subsection{Search sensitivity for ad-hoc waveforms and BBH simulations}
\label{sec:result_sim}
\begin{figure*}[htbp]
\centering
\includegraphics[width=0.99\textwidth]{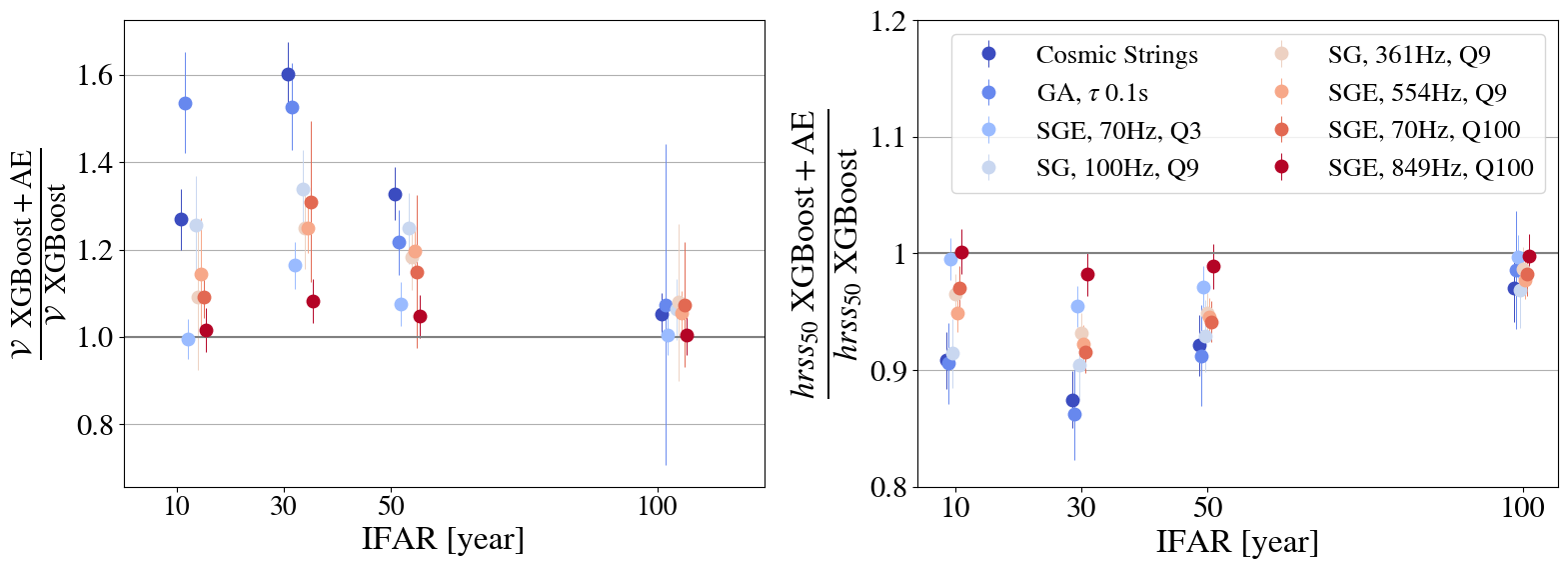}
	\caption{Ratio between the sensitivity volume $\mathcal{V}$ (left) and $h_{rss50}$ (right) obtained including the autoencoder (\texttt{XGBoost + AE}) and without using it (\texttt{XGBoost}) at different IFAR thresholds (10, 30, 50, 100 years). Data points are slightly shifted  around the IFAR thresholds to avoid overlaps. The waveforms are: cosmic strings, Gaussian Pulse (GA) characterized by the duration $\tau$, then Sine Gaussian (SG) characterized by central frequency $f$, and the quality factor $Q$ and White Noise Burst (WNB) with bandwidth $\Delta f$, duration $\tau$ and lower frequency bound $f$. }
	\label{fig:ifar_V}
\end{figure*}

\begin{figure}[htbp]
    \centering
	\includegraphics[width=0.55\columnwidth]{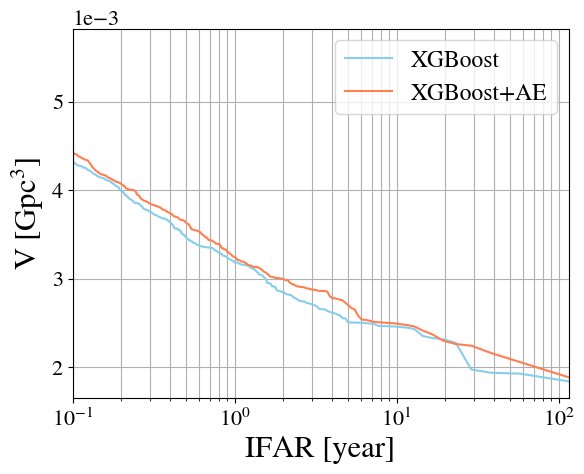}
	\caption{Observed volume at different IFAR for BBH simulations. The volume obtained with the inclusion of the autoencoder (orange) is compared with the model without the autoencoder (blue).}
	\label{fig:roc}
\end{figure}

To evaluate the sensitivity of generic searches, simulated signals are injected into the detector data. 
The  root-squared-sum (rss) strain amplitude of the signal is defined as:
\begin{equation}
    h_{rss} = \sqrt{\int_{-\infty}^{+\infty}[h_{+}^2(t)+h_{\times}^2(t)]dt}
\end{equation}
where $h_{+}$ and $h_{\times}$ are the plus and cross polarization of the GW signal.
Varying the injection amplitude, it is possible to compute the detection efficiency $\epsilon (h_\mathrm{rss})$ as the ratio between the signals detected with an IFAR over a certain threshold and the total amount of injections. From the latter, it is possible to compute the $h_{rss50}$, i.e. the strain amplitude at which $50\%$ of the injections are recovered, which is a typical metric to express the sensitivity in GW bursts searches.\\
A cumulative metric is the sensitivity volume $\mathcal{V}$ \cite{Abadie:2012rq, szczepanczyk2022all}, defined as:
\begin{equation}
    \mathcal{V} = 4 \pi (h_\mathrm{rss,0}r_0)^3 \int_0^\infty \frac{d h_\mathrm{rss}}{h_\mathrm{rss}^4} \epsilon (h_\mathrm{rss}),
    \label{eq:volume}
\end{equation}

where $h_{rss,0}$ is a reference amplitude value at a nominal distance $r_0$. This metric, having a factor $h_{rss}^4$ to the denominator, emphasizes the contribution to the sensitivity of the weaker signals.\\
\begin{figure*}[htbp]
	\includegraphics[width=0.99\textwidth]{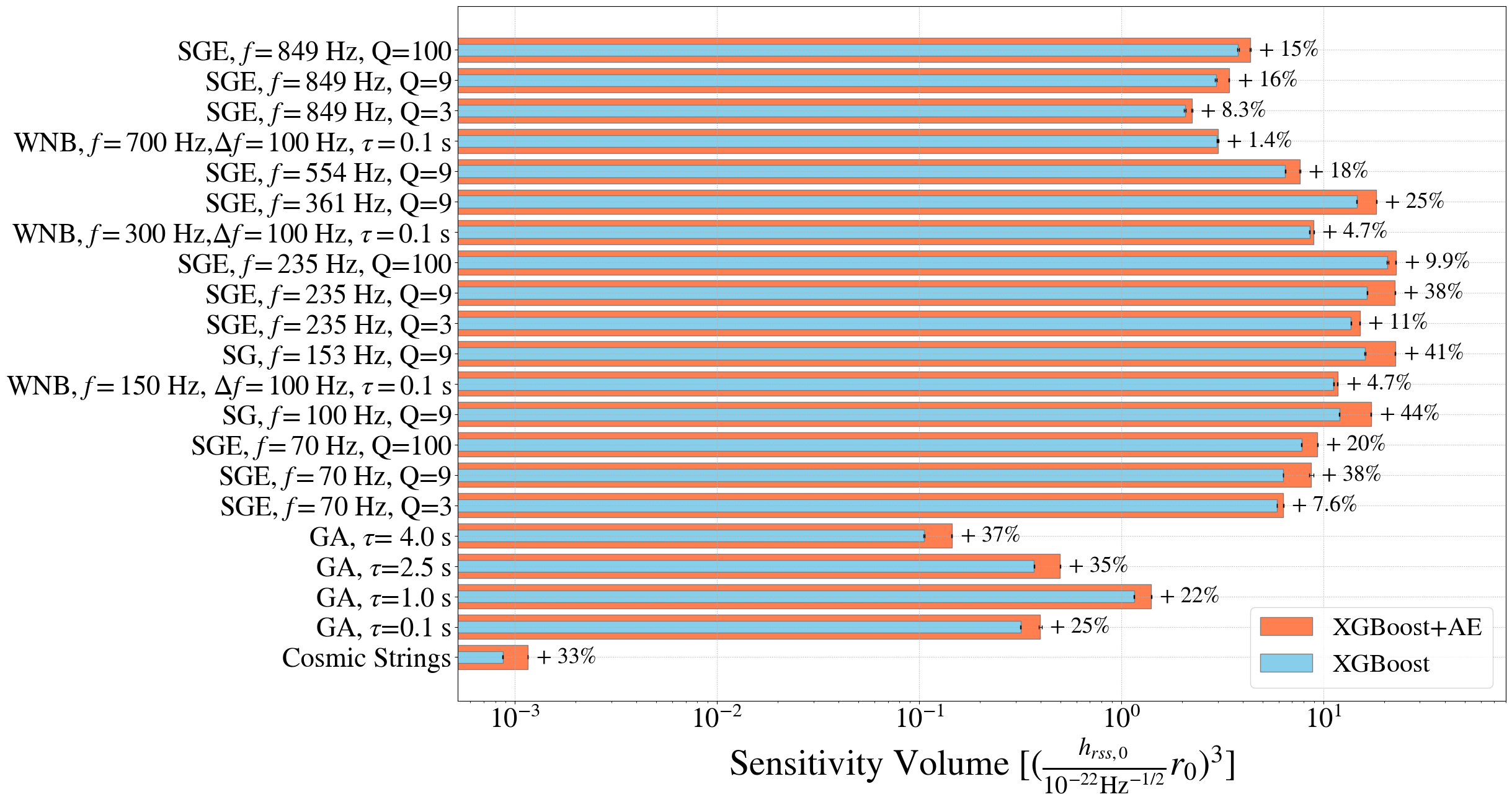}
	\caption{Sensitivity volume obtained with cWB with autoencoder feature included in the XGBoost model (\texttt{XGBoost + AE}) in orange, and without it  (\texttt{XGBoost}) in blue, at IFAR = 50 years. The values next to each bar show the percentage improvement w.r.t to the volume obtained with \texttt{XGBoost} model. From bottom to top, the waveforms are: cosmic strings, Gaussian Pulse (GA) characterized by the duration $\tau$, then ordered according to frequency Sine Gaussian (SG) characterized by central frequency $f$, and the quality factor $Q$ and White Noise Burst (WNB) with bandwidth $\Delta f$, duration $\tau$ and lower frequency bound $f$.}
	\label{fig:volume}
\end{figure*}
Fig.\ref{fig:volume} shows the results for a wide set of ad-hoc waveforms and cosmic strings with an IFAR $>$ 50 years. The sensitivity volume obtained with the inclusion of the autoencoder is higher for all the waveforms tested: for cosmic strings the improvement is $33 \%$ , for GA waveforms between 22 - 37\%, for SG is 8 - 44 \% and for WNB is 1.4 - 4.7 \%. Moreover, we investigate the performance of the two XGBoost models considered over different IFAR thresholds (10, 30, 50, 100 years) in Fig \ref{fig:ifar_V} \footnote{The error bars on the $h_{rss50}$ values are computed from the detection efficiencies at each injected amplitude. The error on the ratio $h_{rss50, \ \texttt{XGBoost + AE}} / h_{rss50, \ \texttt{XGBoost}}$ is propagated assuming the errors on the two terms of the ratio independent or, in other words, neglecting their strong positive correlation. This results in a very conservative choice as the majority of the events is detected by both XGBoost models. 
The error bars on the sensitivity volume are computed from the ones on $h_{rss}$ and by propagating Eq. \ref{eq:volume}}. The improvement in sensitivity volume and in $h_{rss50}$ obtained with the autoencoder statistic 
is remarkable also at lower IFAR thresholds, and it is more evident on the waveforms that have a morphology similar to blip glitches, as GA, cosmic strings and low frequency SG. At IFAR $>100$ years, the search sensitivities obtained by the two models are comparable. This regime corresponds to the background distribution region with $\rho > 13$ (Fig.\ref{fig:bkg}), where there is a single loud glitch which is not affected by the inclusion of the autoencoder. Such background events are rare, and the methodological scope of this work does not justify the computational cost of accumulate more statistic to discuss further this regime. Instead,
we consider relevant the improvement achieved at lower IFAR thresholds, as that is the region where most of the GW signals detected so far lies.\\ 
In addition, we report the sensitivities 
achieved for BBH mergers injections in terms of the observed volume V (Fig. \ref{fig:roc}) \cite{GWTC2}. Given a local merger rate density $R$, this metric is computed counting the number of detections above a certain IFAR threshold $N_{det}$, and considering $N_{det} = \mathrm{VT}R$ where T is the observing time.
The volume obtained including the autoencoder is sightly enhanced for all IFAR thresholds, due to the reduction of the background distribution shown in Fig.\ref{fig:bkg}. 
\section{Conclusions}
In this work, we propose an autoencoder neural network to mitigate the impact of short-duration transient noises, which constitute a major concern for generic gravitational-wave transient (GWT) searches. The neural network is integrated in cWB, a weakly-modeled algorithm widely adopted in the GW community. Over the years, cWB has designed two estimators to recognize short-duration glitches, and recently it has implemented a more efficient separation of GW signals from noises based on the machine-learning algorithm XGBoost \cite{szczepanczyk2022all}. This work constitutes a further step of this development, and shows to improve generic GWT searches in real operating conditions.\\
Here, we focus on blip glitches, one of the most common transient noise family in GW data, whose origin is still unknown.
We include the autoencoder statistic in the XGBoost model, and we perform injections of ad-hoc waveforms, cosmic strings and BBH simulations in GW data. We report the sensitivity achieved both in terms of sensitivity volume and $h_{rss50}$. The inclusion of the autoencoder statistic enhances ad-hoc waveforms and cosmic strings at different IFAR thresholds, in particular the most evident enhancement is achieved for the waveforms which have a morphology similar to blip glitches, as short-duration gaussian pulses, sine gaussians and cosmic strings. The search sensitivity for BBH simulations is also slightly enhanced by the addition of the autoencoder statistic.\\
Here, the methodology is applied to the LIGO detector network, but it could be easily extended to multiple GW detectors. In addition, the autoencoder statistic could be exploited by other signal-noise classification procedures, as the one based on Gaussian Mixture Model recently proposed for cWB \cite{gmmcwb, gmmcwb2022}.\\
With respect to previous deployed methods, as the $Q_{veto}$ parameters, the autoencoder neural network is able to learn also different transient noise classes, if present in the training dataset. This flexibility will be highly valuable as new glitch classes appear in the future GW observing periods.

\ack
This material is based upon work supported by NSF's LIGO Laboratory which is a major facility fully funded by the National Science Foundation. The authors are grateful for computational resources provided by the LIGO Laboratory and supported
by National Science Foundation Grants PHY-0757058
and PHY-0823459. This research has made use of data, software and/or web tools obtained from the Gravitational Wave Open Science Center, a service of LIGO Laboratory, the LIGO Scientific Collaboration and the Virgo Collaboration

\section*{Appendix}
\renewcommand{\thesubsection}{\Alph{subsection}}
\subsection{Qveto parameters}
\label{sec:Qveto}
\begin{figure}[ht]
    \centering\includegraphics[width=0.65\columnwidth]{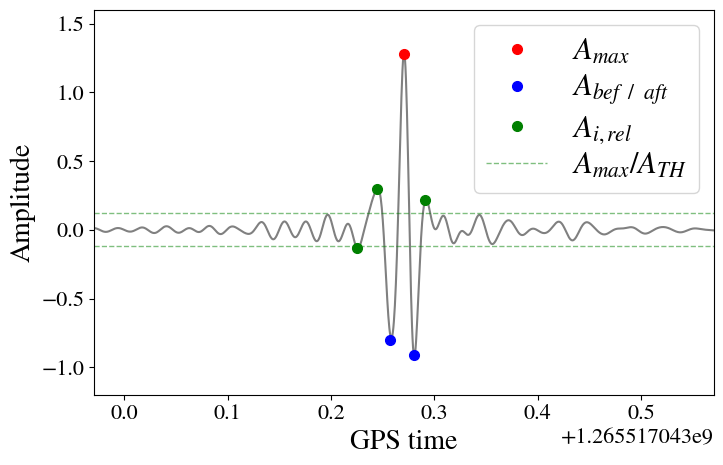}
	\caption{Reconstructed waveform of a transient noise passing in cWB in LIGO Hanford. The coloured dots show the amplitudes used to estimate the $Qveto_0$ according to Eq. \ref{eq:qveto0}. The green lines correspond to the threshold $\pm A_{max}/A_{TH}$ over which $A_{i,rel}$ amplitudes are chosen.}
	\label{fig:Qveto0}
\end{figure}

cWB calculates several summary statistics to characterize the detected events. Here we focus on two of them, called $Qveto$, introduced to pinpoint short-duration glitches and reviewed here for the first time\footnote{See also in the cWB documentation:\\ https://gwburst.gitlab.io/documentation/latest/html/faq.html$\#$the-qveto }.\\
The first, $Qveto_0$, estimates the ratio between the energies far and near the maximum peak of the signal. 
More precisely, once computed the absolute maximum amplitude $A_{max}$, $N_{TH}$ samples  before and after 
the maximum peak are selected. Their amplitudes are indicated with $A_{i, bef}$ and $A_{i,aft}$. Further relative amplitudes $A_{i,rel}$ are selected if $|A_{i,rel}|>|A_{max}|/A_{TH}$. Fig. \ref{fig:Qveto0} shows the cWB reconstructed waveform of a transient noise with the relevant amplitudes for the estimation of the $Qveto_0$ highlighted. $Qveto_0$ is then defined as:
\begin{equation}
    Qveto_0 = \frac{\sum{A_{i,rel}}^2}{A_{max}^2 + \sum{A_{i,bef/aft}^2}}
\label{eq:qveto0}
\end{equation}
The lower the $Qveto_0$, the strongest the peak amplitude compared to the surrounding fluctuations, suggesting a blip-like morphology. $N_{TH}$ and $A_{TH}$ are hard thresholds which are empirically selected looking at glitches reconstructed waveforms. Default values are $N_{TH} = 1$, and $A_{TH} = 7.6$. This procedure is applied to each detector independently, then the minimum value is selected.
The second parameter, $Qveto_1$, models approximately the reconstructed waveform 
with a CosGaussian function:
\begin{equation}
    \mathrm{CosGaussian}(w,Q) = \mathrm{cos}(\omega t) *\mathrm{exp}\bigg[\frac{(- \omega t)^2}{2 \ {Qveto_1}^2} \bigg]
\end{equation}
and the $Qveto_1$ factor is estimated as:
\begin{equation}
    Qveto_1 = \sqrt{\frac{-\pi^2}{2 \mathrm{log}( R)}},  \ \mathrm{with} \
    R = \frac{A_{bef}+A_{aft}}{2A_{max}}
\end{equation}
where $A_{bef}$ ($A_{aft}$) is the absolute value of the maximum peak before (after) the main peak. The $Qveto_1$ is computed for each detector and weighted according to the SNR square of the detected event in each detector.\\ 
On LVK publications based on the data from the first, second and third observing runs (O1-O2-O3) \cite{abbott2017all,abbott2019all,abbott2021all}, $Qveto$ parameters were employed in cWB to split the detected events into multiples bins: a 'clean' bin and one or more 'dirty' bins. Dirty bins were populated by short-duration blip-like events, while the clean bin contains only triggers with $Qveto$ parameters above a certain threshold (as an example for the O3 generic GW burst search \cite{O3allskyshort} $Qveto_1 <3.4$ for the first dirty bin, $Qveto_1 <3.4$ and $Qveto_0 = 0$ for the second dirty bin, and $Qveto_1 >3.4$ for the clean one). Then, each event was ranked in each bin independently, leading to the introduction of a trial factor equal to the number of bins. Thus, $Qveto$ parameters played a key role during the signal-noise separation process. \\
However, this procedure depends on hard thresholds that had to be tuned manually according to the performance on some set of simulations, and it cannot be generalized to different transient noise morphology that may arise in next observing runs. For these reasons, it has been substituted with a machine-learning based algorithm \cite{mishra2021optimization, mishra2022search, szczepanczyk2022all} which learns the population of the signal and of the noise from a list of cWB summary statistics.
Among the others, this list includes the $Qveto$ parameters through the following definitions:
\begin{equation}
    Q_a = \sqrt{Qveto_0} \ , \\
   Q_p = \frac{Qveto_1}{2 \sqrt{\mathrm{log}_{10}(\mathrm{min}(200, E_{c}))}}
\end{equation}
where $E_{c}$ 
is the coherent energy in the detector network (Section \ref{sec:cWB}). 
\subsection{More details on the autoencoder neural network}
\label{appendix}
We report here more details on the methodology implemented. The algorithm is based on deep-learning, a subset of machine-learning, in which a task is learned from a large amount of data. Deep-learning methods consist of a network, i.e. a sequence of layers of algorithms, each of which extrapolate information of the data it is applied to. These networks are usually referred as \textit{neural network} because they are inspired by the human brain processing. The neural network proposed in this works consists of two parts, an encoder that compress the input into a lower dimensional space and a decoder that returns the compressed representation into the original shape. Both the encoder and the decoder are made of multiple layers. The neural networks with this structure are called \textit{autoencoders}. The first layer learns a representation of the input data, and the subsequent layers receive the representation learned by the preceding layers, revealing more and more complex features. The main layers used are briefly introduced below and summarized in Table \ref{tab:arch}. A review can be found in Ref. \cite{alzubaidi2021review}. 
\begin{itemize}
\item \textit{Convolutional layer}:  
it computes the convolution between the input $x$ and the filters, or \textit{kernels}. Multiple kernels are present in each convolutional layers, and have dimension equal to $(m \times k)$, being $m$ the length of the stride and $k$ the number of kernels applied. The convolutional layer calculates the dot product between the input and the weight $W^k$ and transmits the result of the multiplication to an activation function $f$. The output of the convolutional layer is:
\begin{equation}
    h_{W,b}=f(W^k x+b)
    \label{eq:NN}
\end{equation}
where $b$ is a bias term, typically equal to 1. Here, $f$ is a ReLU function \cite{agarap2018deep}, which converts the input to positive numbers:
\begin{equation}
f(z) = \mathrm{max} ( 0, z )
\end{equation}
The kernels are assigned randomly at the beginning of the training process and are updated minimizing the error function (Eq. \ref{eq:loss}). The filter length $m$ is equal to 3 and slides over the entire input. Usually, multiple filters are used to acquire more complex kinds of features. 
\item \textit{Max Pooling layer}: after a convolutional layer typically there is a pooling layer, that down-samples the convolutional output $h_{W,b}$ picking the maximum value over a spatial window considered. In this case, we have a window equal to 2, meaning that we take the maximum values between two adjacent values. The combination of convolutional layers and max pooling layers is repeated multiple times to extract the most relevant features.
\item \textit{Dense layer}: 
it consists of several basic units, called \textit{neurons}, in which a weight is applied as in Eq. \ref{eq:NN}. Each neuron is fully connected to all neurons of the previous layers. In this autoencoder architecture a dense layer is used to compress the output of the decoder layers to the desired latent space dimension, equal to 200 in this case.
\item \textit{Up Sampling layer}: opposite to Max Pooling layer up-sample the representation repeating the data by 2.
\item \textit{Flatten} and \textit{Reshape}: the first flatten the inputs from a shape $x^{a,b}$ to $x^{a \times b}$, the second reshapes the inputs into the given shape.
\end{itemize}
The main settings used to train the network are: the epochs, i.e. the number of iterations over the entire training dataset, sets to 75, the bach size, i.e. the number of training samples per weights update, that is 16, and the optimization algorithm which updates the network weights minimizing the loss function that is ADAM \cite{kingma2014adam}.
\begin{center}
\begin{table}
\centering
\small
\begin{tabular}{c c} 
 \hline
 Layer & Output Shape (length, dimension)\\ [0.5ex]
 \hline\hline
 \multicolumn{2}{c}{\textbf{Encoder}}\\
 Input & (416,1) \\ 
 Convolutional & (416,128) \\
 Max Pooling & (208,128) \\
 Convolutional & (208,16) \\
 Max Pooling & (104, 16) \\
 Convolutional & (52, 16) \\
 Flatten & (832) \\
 Dense & (200) \\
 \multicolumn{2}{c}{\textbf{Decoder}}\\
 Dense  & (832) \\
 Reshape & (52, 16) \\
 Convolutional & (52,16) \\
 Up Pooling & (104,16) \\
 Convolutional & (104,16) \\
 Up Pooling & (208,128) \\
 Convolutional & (208,128) \\
 Up Pooling & (416,128) \\
 Convolutional & (416,1) \\
 \hline
 \hline
\end{tabular}\caption{\label{tab:arch} Autoencoder neural network architecture. Each line represents a layer of algorithm. On the right, there is the output shape of each layer, which is also the input shape for the subsequent layer. For example: the input data is a time-series with 416 data points. Then, a convolutional layer with $k$=128 filters is applied. The time-series sample rate is 2048 Hz. }
\end{table}
\end{center}
\section*{References}
\bibliography{iopart-num}

\end{document}